\documentclass[final,3p,times,sort&compress]{elsarticle}

\usepackage{setspace,rotating,url}
\usepackage{amsmath,amssymb,amsfonts}
\usepackage{algorithmic}
\usepackage{graphicx}
\usepackage{textcomp}
\usepackage{epstopdf}
\usepackage{caption}
\usepackage{subfigure}
\usepackage{multirow}
\usepackage{array}
\usepackage{tabularx}
\usepackage{booktabs}
\usepackage{colortbl,color}
\usepackage{natbib}

\usepackage{amssymb}

\begin{document}

\begin{frontmatter}



\title{Exploring the effect of social media and spatial characteristics during the COVID-19 pandemic in China }

\author[1]{Xiu-Xiu Zhan}
\author[1]{Kaiyue Zhang}
\author[1]{Lun Ge}
\author[2]{Junming Huang}
\author[5]{Zinan Zhang}
\author[3]{Lu Wei}
\author[4]{Gui-Quan Sun}
\author[1]{Chuang Liu}
\author[3]{Zi-Ke Zhang} \ead{zkz@zju.edu.cn}
\address[1]{Research Center for Complexity Sciences, Hangzhou Normal University, Hangzhou 311121, PR China}
\address[2]{Paul and Marcia Wythes Center on Contemporary China
and Department of Sociology, Princeton University}
\address[3]{College of Media and International Culture, Zhejiang University, Hangzhou 310058, China}
\address[4]{Department of Mathematics, North University of China, Taiyuan, Shanxi 030051, China}
\address[5]{School of Public Finance and Taxation, Zhejiang University of Finance and Economics, Hangzhou 310018, China}

\begin{abstract}
The declaration of \textit{COVID-19} as a pandemic has largely amplified the spread of related information on social medium, such as Twitter, Facebook and WeChat. Unlike the previous studies which focused on how to detect the misinformation or fake news related to \textit{COVID-19}, we investigate how the disease and information co-evolve in the population. We focus on \textit{COVID-19} and its information during the period
when the disease was widely spread in China, i.e.,
from January 25th to March 24th, 2020. We first explore how the disease and information co-evolve via the spatial analysis of the two spreading processes.
We visualize the geo-location of both disease and information at the province level and find that disease is more geo-localized compared to information. We find high correlation between the disease and information data, and also people care about the spread only when it comes to their neighborhood. Regard to the content of the information, we find that positive messages are more negatively correlated with the disease compared to negative and neutral messages. Additionally, we introduce machine learning algorithms, i.e., linear regression and random forest, to further predict the number of infected using different characteristics, such as disease spatial related and information-related characteristics. We obtain that the disease spatial related characteristics of nearby cities can help to improve the prediction accuracy. Meanwhile, information-related characteristics can also help to improve the prediction performance, but with a delay, i.e., the improvement comes from using, for instance, the number of messages 10 days ago, for disease prediction. The methodology proposed in this paper may shed light on new clues of
emerging infections prediction.

\end{abstract}




\end{frontmatter}

\section{Introduction}\label{sec:introduction}

\makeatletter
\newcommand{\rmnum}[1]{\romannumeral #1}
\newcommand{\Rmnum}[1]{\expandafter\@slowromancap\romannumeral #1@}
\makeatother
\textit{COVID-19} has become a world-wide pandemic since it was first reported in the late 2019 in Wuhan, China. It has coursed more than 90 million infected and 2 million deaths all over the world until January 2021.
As a new contagious disease, it is the first time that the internet and social media are used as tools to inform the pandemic, share knowledge, keep people connected at the quarantine
 time, etc.  Until now, the treatment of \textit{COVID-19} is rather limited since no vaccine is approved to be effective. Thus, one may rely more on the information s/he obtains online to guide her/him to take actions, such as wear masks, wash hands and keep distance, against the disease~\citep{betsch2020social,cheng2020wearing, zhang2021multiplex, jia2020population,liu2020covid,zhong2021mental,choudrie2021machine}.

 Disease information appears and co-evolves with the disease when a disease starts to spread in the population. The development of internet has changed the way of information transmission from offline to online, which largely accelerate the transmission speed and also dramatically increase the outbreak size of the information. For example, the maximum daily number of individuals talking about H7N9 in 2013 on the largest micro-blogging system in China, \textit{Sina Weibo}, was almost 90 million~\citep{zhan2018coupling,zhang2016dynamics, liu2020computational,wang2019coevolution}. For \textit{COVID-19}, we have collected 163.1 million messages (including 154.8 geotagged messages) on Chinese online social platforms from January 25th, 2020 to March 24th, 2020, with 2.58 million messages per day on average. Among the information spreads on the social platforms, one may find reliable information that could help to adopt norms or behaviors which can inhibit the disease. However, there are also low-credibility or false information that may be harmful to public's health both physically and mentally and also may cause panic in the population. The information related to \textit{COVID-19} has been recognized as infodemic by WHO, as it contains overabundance and questionable information~\citep{zarocostas2020fight}. The availability of the information data online has motivated researchers to work on interesting research problems~\citep{hu2021infectivity, sun2021transmission,poirier2020real,zhang2020evolving,sun2020transmission,prasse2020network}. For example, Gallotti et al.~\citep{gallotti2020assessing} have worked on 100 million Twitter messages to investigate the risks of infodemics in different countries. They proposed an infodemic risk index to capture unreliable news across countries. Cinelli et al.\citep{cinelli2020covid} analyzed the information diffusion data of \textit{COVID-19} on platforms such as Twitter, Instagram, YouTube, Reddit and Grab. They modeled the information diffusion with epidemic model on different social platforms to obtain the basic reproduction number. There are also papers working on how to detect the misinformation and social bots related to \textit{COVID-19}~\citep{cresci2020decade,roitero2020covid,tagliabue2020pandemic,liu2021mobility, aleta2020data}.

 Despite the effort on the analysis and modeling of \textit{COVID-19} and its information, the coupling effect between these two dynamical processes remains unknown. In this work, we research the coupling dynamics between the spread of \textit{COVID-19} and its information. We collect the information data from Qingbo Bigdata (\url{www.gsdata.cn}) with specific key Chinese words related to \textit{COVID-19}. The data contains most of the well-known social medium in China and is representative regard to the public response to the disease.
 We analyze the geo-location data of both disease and information, and explore the correlation between these two evolutionary processes. We find that the province-level disease and information data are highly correlated. For the disease, the provinces that are close to Hubei are more infected than the others. However, people in the more developed provinces, such as Beijing, Guangdong and Shanghai, tend to discuss about \textit{COVID-19} online more than the other provinces. This indicates that disease is more spatially localized compared to information. Based on the analysis of \textit{COVID-19} and its information data, we further explore whether information can help to improve the prediction of \textit{COVID-19} by machine learning models, i.e., linear regression and random forest~
\citep{achterberg2020comparing}.

 The rest of the paper is organized as follows. In Section \textbf{2}, we give detailed analysis of the information and disease data at the province level. Based on the correlation analysis in the above section of disease and its information, we further use machine learning algorithms, i.e., linear regression and random forest, to predict how many people will be infected in the future in Section \textbf{3}. We note that we predict the number of infected in each city. Additionally, we also analyze which characteristics are more important for the disease prediction in this section. We conclude the paper in Section \textbf{4}.
 The detailed descriptions of the algorithms and data are given in
Section \textbf{5}.

\section{Spatial analysis of COVID-19 and its information}

\begin{figure*}[!ht]
\centering
	\includegraphics[width=13cm]{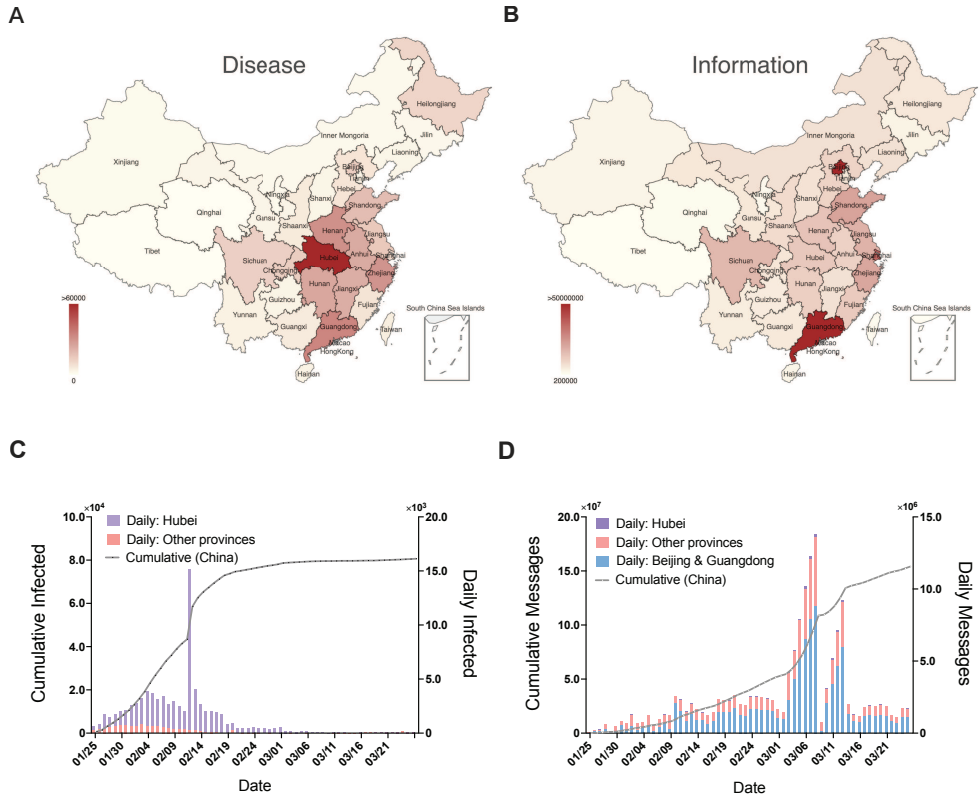}
    \caption{Visualization of (A) the total number of infected cases and (B) the total number of messages in each province of China.
    (C) We show the daily number of infected in Hubei (purple bar), the daily number of infected in the other provinces except Hubei (light red bar) and the cumulative number of infected cases in China (black dash line). (D) We show daily number of messages in Hubei (purple bar), daily number of messages in the other provinces except Hubei (light red bar), daily number of messages in Beijing and Guangdong (blue bar) and the cumulative number of messages in China (black dash line).}
    \label{fig:visualization-globally}
\end{figure*}

In Figure~\ref{fig:visualization-globally}A and~\ref{fig:visualization-globally}B, we visualize the total number of infected and the total number of messages regarding to \textit{COVID-19} in each province. The exact quantities of the two variables are given in Table~1 in the Appendix. We find that the infected cases of \textit{COVID-19} are mostly distributed in the provinces that are geographically close to Hubei province.
The top $10$ provinces with the most infected individuals are Hubei, Guangdong, Henan, Zhejiang, Hunan, Anhui, Jiangxi, Shandong, Jiangsu and Chongqing. The top $10$ provinces with the most number of messages are Beijing, Guangdong, Shanghai, Shandong, Zhejiang, Jiangsu, Sichuan, Henan, Hubei and Fujian. Thus, 6 provinces, i.e., Hubei, Guangdong, Henan, Zhejiang, Shandong, Jiangsu, are not only in the top $10$ most infected but also in the top $10$ most informed. Hubei is the original place where \textit{COVID-19} was first reported and is also the place with $82.8\%$ of infected individuals all over China.
However, it seems people from the most developed provinces, such as Beijing, Guangdong and Shanghai are more likely to talk about \textit{COVID-19} online compared to Hubei. It is validated by the fact that the more developed provinces tend to have more messages about \textit{COVID-19} in Figure~S6A and S6B in the Appendix. In the figure, we show the number of messages is highly correlated with the GDP and GDP per capita in every province, with the Pearson correlation coefficients equal to $0.77$ $(p=3.8 \times 10^{-7})$ and $0.71$ $(p=8.9 \times 10^{-6})$, respectively.

\begin{figure*}[!ht]
\centering
	\includegraphics[width=13cm]{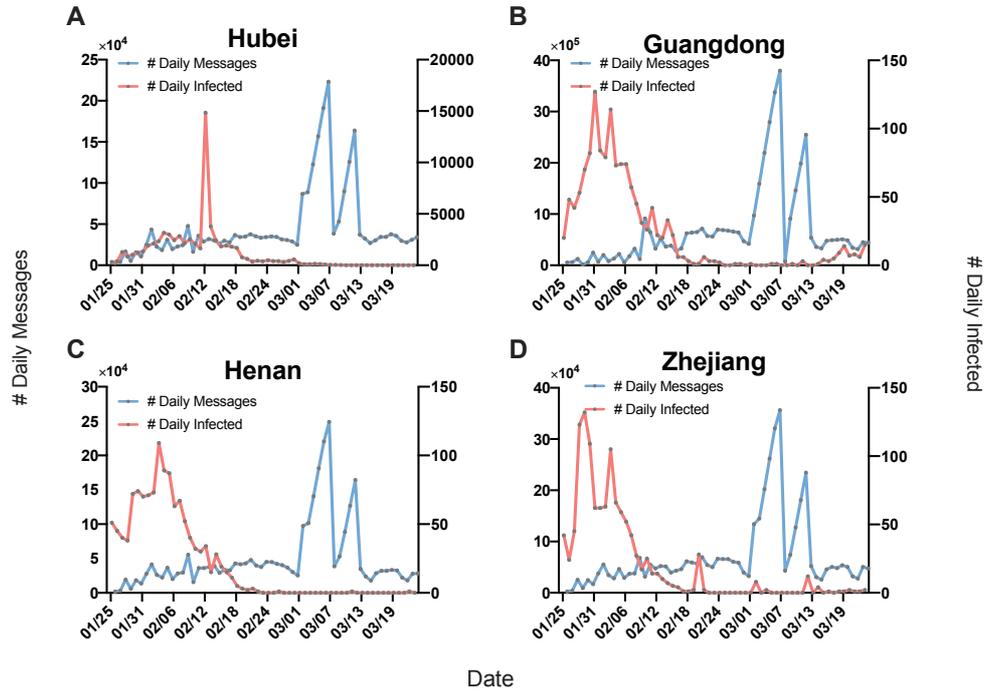}
    \caption{Daily number of infected (red curve) and messages (blue curve) in each province. We show the top 4 provinces (i.e., Hubei, Guangdong, Henan, Zhejiang) that have most infected individuals. The daily number of infected (messages) of the other provinces are given in Figure S4  in the Appendix.}
    \label{fig:province-daily-infected}
\end{figure*}

\begin{figure*}[!ht]
\centering
    \includegraphics[width=6cm]{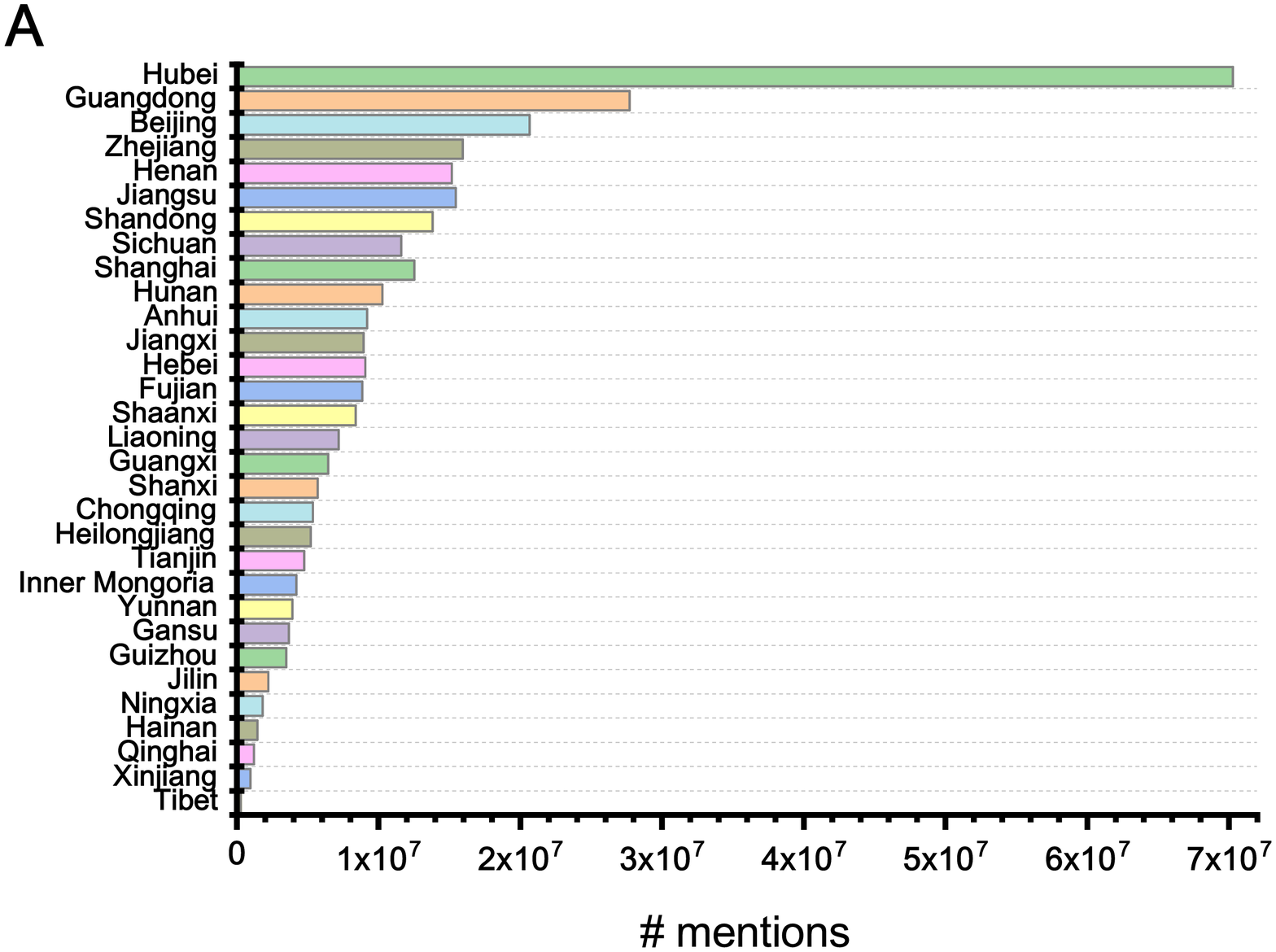}
    \includegraphics[width=6cm]{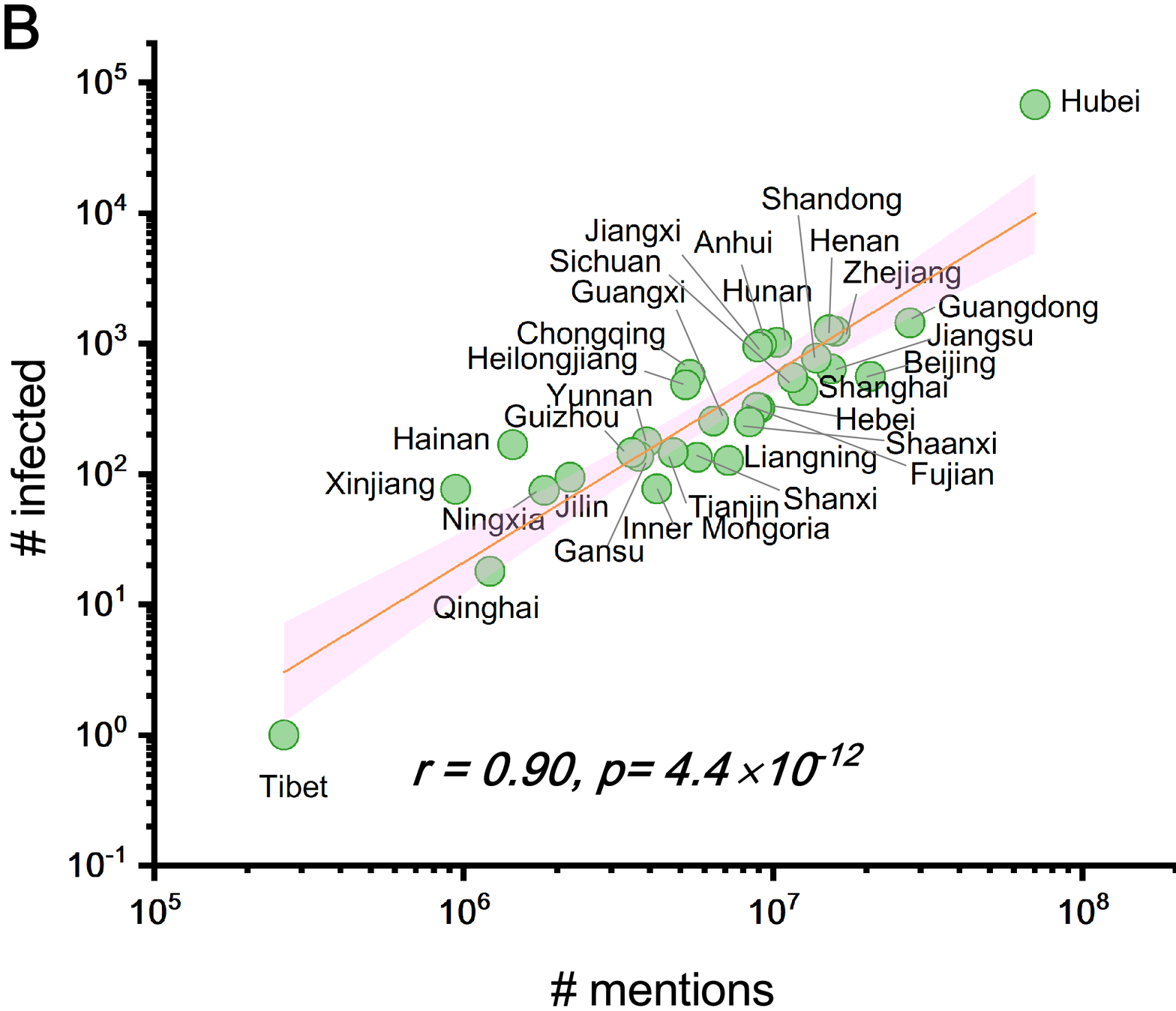}
    \caption{(A) The number of times a province is mentioned in all the messages. (B) The correlation between the number of mentioned times and the number of infected at the province level.}
    \label{fig:mention}
\end{figure*}
In Figure~\ref{fig:visualization-globally}C (D), we further show the daily increase of infected (messages) and cumulative number of infected (messages) globally in China. To be specific,
 the disease spreads fast during January and February and becomes stable in March, with the most cases from Hubei province (shown by the black dash curve in Figure~\ref{fig:visualization-globally}C). A significant spike in the number of infected cases on 13th of February is because the health officials essentially broadened the definition of what could be counted as an infected case.
 However, the number of messages continuously increases during the research period, with Beijing and Guangdong dominating the diffusion of information.
 The spreading pattern of \textit{COVID-19} and its information shows that despite the coevolution of these two spreading processes, the peak of \textit{COVID-19} information is much later than that of \textit{COVID-19} in China. As a matter of fact, when a new contagious disease starts to spread in the population, people may not take it serious at the beginning as they lack knowledge about the disease. However, when people are aware of the high infection and fatality rate of the disease, they will pay more attention and start to talk both online and offline. Thus, information starts to break out in the population. The time lag between the outbreak of disease and information has also been found in \citep{zhan2018coupling}, where we studied the disease and information spreading for \textit{H7N9} and \textit{Dengue}.
Additionally, we further confirmed the time lag between these two spreading processes by giving the daily increase of disease and information in Figure~\ref{fig:province-daily-infected} at the province level. In Figure~\ref{fig:province-daily-infected}, we only show the daily increase for the top 4 most infected provinces, the daily increase for the other provinces are shown in Figure~S4 in the Appendix.  For most of the provinces, the number of infected increased before 21st, February. The daily number of messages shows similar increasing pattern in the provinces, with the peak around 6th of March. If we take a look of the messages' content, we find that the content may mention the names of the cities (or provinces). In Figure~\ref{fig:mention}A, we show the number of times a province is mentioned and find that Hubei is the most mentioned province. The number of times that Hubei is mentioned is much higher than the second most mentioned province, i.e., Guangdong. We also visualize how the top 5 most mentioned provinces are mentioned by the other provinces in Figure~S5 in the Appendix. We find that the top 5 most mentioned provinces are more mentioned by people from provinces, such as Beijing and Guangdong.
Furthermore, the number of infected is highly correlated with the number of times a province is mentioned, as shown in Figure~\ref{fig:mention}B. The Pearson correlation coefficient  in the log-log scale between them is as high as $0.90$ $(p=4.4 \times 10^{-12})$.

In Figure~\ref{fig:province-daily-infected} and S4 in the Appendix, the daily number of messages in different provinces shows similar trend of peak. However, the evolution of information and the total number of messages are still different across provinces. Therefore, we are motivated to explore what kind of factors are correlated with the quantity of messages posted in every province.
We show how the number of messages in every province is correlated with the following variables, i.e., province-wise population, the distance from Hubei to the corresponding province, GDP, GDP per capita, electricity consumption and highway transport volume in each province in Figure~\ref{informationPCC} (Figure~S6 in the Appendix) .
 For simplification, we denote the total number of messages in every province as information volume.
The distances between Hubei and other provinces are computed according to their capitals' geographic location. We show the Pearson correlation between the variables. But in different figures, we may use linear or log scale of the variable for the correlation calculation. Taking population as an example, we compute the Pearson correlation coefficient (PCC) between the the information volume list and population list of all the provinces in the log-log scale. The result
is shown in Figure~\ref{informationPCC}A. The PCC between the information volume and population size is $r=0.57$ $(p=7.7 \times 10^{-4})$, indicating provinces with higher population size tend to have larger volume of information. The PCC between the information volume and the province distance to Hubei is negative, which equals to $r=-0.44$ $(p=1.4 \times 10^{-2})$. Actually, if we take a look of Figure~\ref{informationPCC}D, the number of infected in the province-wise level is also negatively correlated with the province distance to Hubei. This means that the less population will be infected if the province is further away from Hubei.
As people in the provinces which are further away from Hubei feel less threatened by the disease, resulting in a smaller information volume in the corresponding province. That is to say, people care about the spread only when it spatially comes to their neighborhood. We investigate the correlation between the number of infected and information volume in province-wise level in Figure~\ref{informationPCC}C, which shows a high correlation ($r=0.58, p=6.3 \times 10^{-4}$). Figure~S7A-D in the Appendix indicate the number of infected is also highly correlated with economic
characteristics such as GDP, GDP per capita, electricity consumption and highway transport volume. Since \textit{COVID-19} started to spread in the population, the only effective way to decrease the infected number is lockdown. However, lockdown has affected the normal economic activities~\citep{nishi2020network}. The high correlation between \textit{COVID-19} and the economic
characteristics shown in Figure~S7A-D indicates that sustaining the economic activities and inhibiting the disease spreading is more urgent in developed areas.

\begin{figure*}[!ht]
\centering
	\includegraphics[width=13cm]{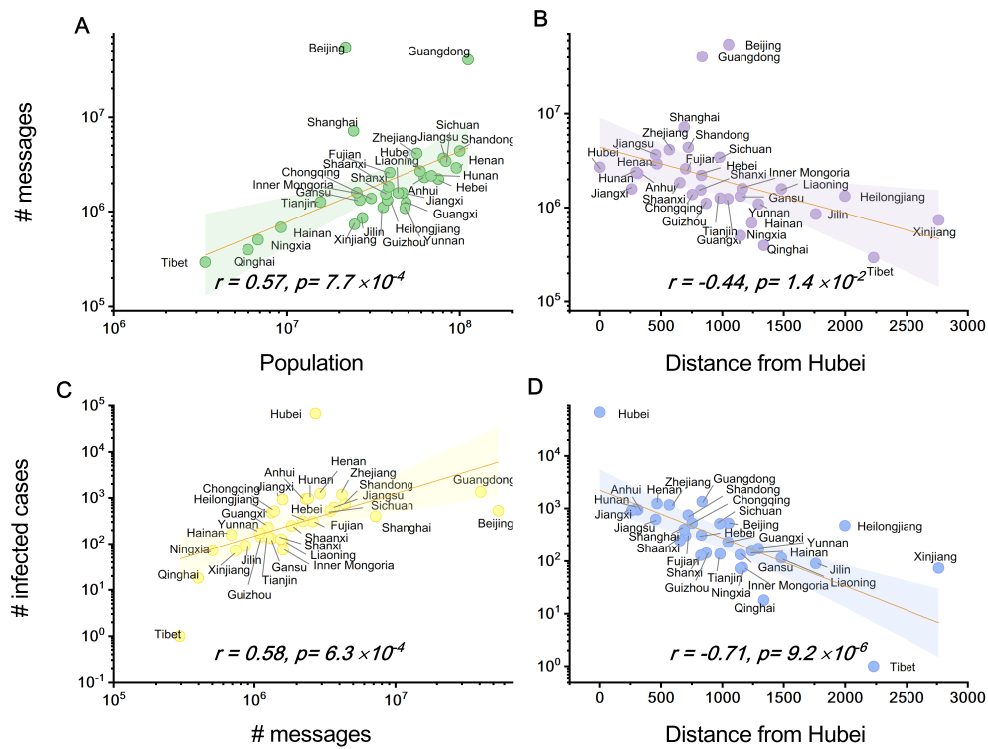}
    \caption{Correlation between the total number of messages in each province and (A) population in each province; (B) the distance from Hubei to each province. Correlation between the total number of infected in each province and (C) the total number of messages in each province; (D) the distance from Hubei to each province.}
    \label{informationPCC}
\end{figure*}

\section{City-wise Epidemic Prediction}

Armed with the above analysis about the relation between COVID-19 and its information, we explore whether the diffusion of information about the disease can help to predict the disease. For the disease and information, we have the data of every single city. Therefore, we will focus on the prediction of disease in the city level in this section.
We use linear regression (LR) and random forest (RF) algorithms as prediction models. The details of these two algorithms are given in Section~\textbf{Method and Data Description}. The features derived from the disease as well as its information are used as features for prediction.

Let us formally present our method. We consider to predict the number of infected in each day of $322$ cities in China. The duration of \textit{COVID-19} data we use in our prediction model is from January 25th to March 9th, 2020. Thus we use the first 34 days as the training data and the remaining 10 days as the test data. As a baseline model, we first introduce how to use the historical
disease data to predict the number of infected. Then, we further add the historical data of the disease information into the baseline model to explore whether the features derived from disease information is helpful for improving the prediction accuracy.
\subsection{Evaluation Metric}
We use mean absolute error (\textit{MAE}) to evaluate the performance of the prediction model. For instance, if the real number of infected in each city at time $t$ is given by $\{I_{t}^{1}, I_{t}^{2}, \cdots, I_{t}^{N}\}$, where $N$ is the number of cities.  Variable $I_{t}^{j}$ means the number of infected of city $j$ at time $t$. The predicted values are given by $\{\hat{I_{t}^{1}}, \hat{I_{t}^{2}}, \cdots, \hat{I_{t}^{N}}\}$.  Then \textit{MAE} at time $t$ is given as follows:

\begin{equation}
\centering
MAE_t = \frac{1}{N} \sum_{j=1}^{N}|I_{t}^{j}-\hat{I_{t}^{j}}|
\end{equation}

 If we consider the prediction of ten days, we can use the average \textit{MAE} values over the ten days, denoted as \textit{<MAE>}, to evaluate the prediction performance.
The small value of \textit{MAE} (or \textit{<MAE>}) means the predictive model is more accurate in predicting the number of infected, and vice verse. We note that we have 322 cities with nonzero infected cases. Thus in the following analysis, we use $N=322$.

\subsection{Prediction based on disease historical data}
The historical data of the disease is the number of infected at each day for every city. The number of infected in city $j$ at day $t$ is given by $I_{t}^{j}$.
For every city, we show how the current number of infected is correlated with the historical number of infected in that city by computing the auto-correlation between the time series. We use $\tau$ to represent the time lag.  For a given city $j$, suppose the number of infected time series is given by $I^{j}=\{I_{1}^{j}, I_{2}^{j}, \cdots, I_{T}^{j}\}$, the lag $\tau$ auto-correlation function $r_{\tau}^{j}$ is defined as
\begin{equation}
\centering
r_{\tau}^{j} = \sum_{t=1}^{T-\tau}\frac{(I_{t}^{j}-\bar{I^{j}})-(I_{t+\tau}^{j}-\bar{I^{j}})}{\sum_{t=1}^{T}(I_{t}^{j}-\bar{I^{j}})^2},
\end{equation}
where $\bar{I^{j}}$ is the average number of infected in city $j$. The average lag $\tau$ auto-correlation $r_{\tau}$ for all the cities is defined as \begin{equation}
\centering
r_{\tau}=\sum_{j=1}^{322}r_{\tau}^{j}
\end{equation}
\begin{figure*}[!ht]
\centering
	\includegraphics[width=12cm]{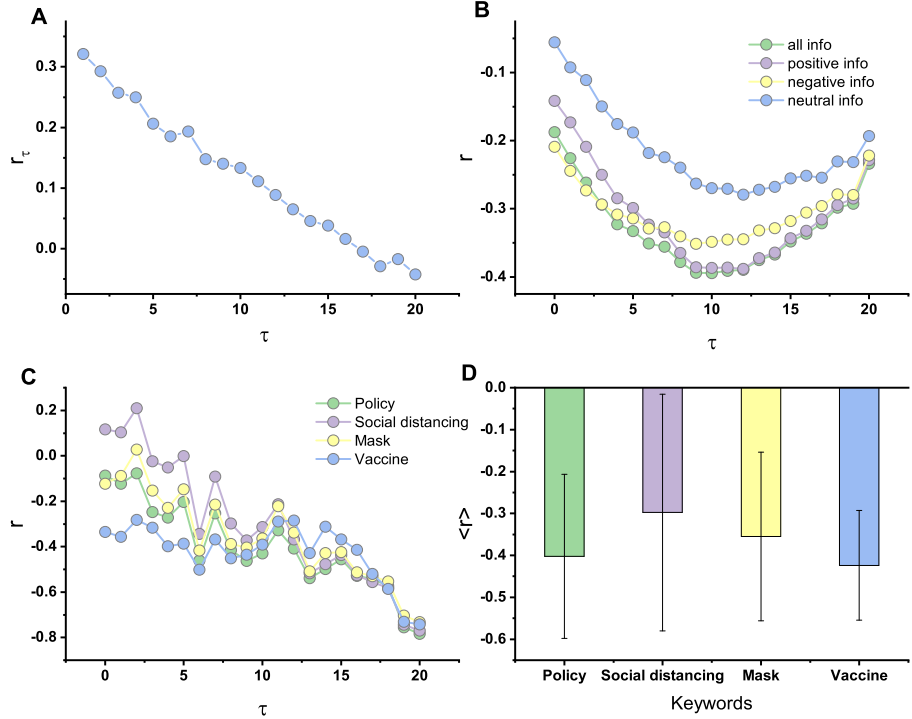}
	\caption{(A) The change of the average lag $\tau$ ($\tau \in [1, 20]$) auto-correlation $r_{\tau}$, $r_{\tau}$ is given by Equation~(3) for the infected time series. (B) Pearson correlation between the infected time series and information time series with different delay values $\tau$ ($\tau \in [1, 20]$), different colors indicate we use different types of information, i.e., all the messages (green), positive messages (purple), negative messages (yellow) and neutral messages (blue). (C)  Pearson correlation between the infected time series and words frequency time series with different delay values $\tau$ ($\tau \in [1, 20]$), different colors indicate we use different types of words, i.e., Policy (green), Social Distancing (purple), Mask (yellow) and Vaccine (blue). (D) The average Pearson correlation across different values of $\tau$ for the four words given in C. }
    \label{lag tau autocorr}
\end{figure*}
We show how the average lag $\tau$ auto-correlation $r_{\tau}$ is changing with lag $\tau$ in Figure~\ref{lag tau autocorr}A, where $\tau \in [1, 20]$. The auto-correlation coefficient decays with the increase of $\tau$, which means the number of infected is more related to the recent number of infected. When $\tau=1$, we achieve the highest auto-correlation coefficient.

\begin{figure*}[!ht]
\centering
	\includegraphics[width=12cm]{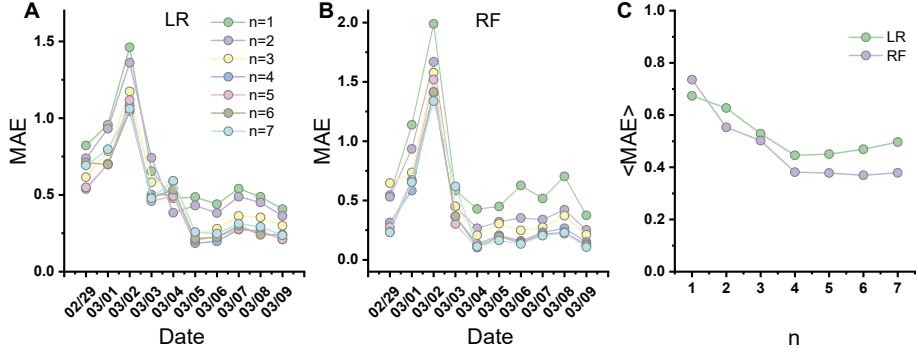}
	\caption{Prediction based on using the target city's historical disease data as features. (A) Daily prediction performance (\textit{MAE}) for linear regression; (B) daily prediction performance (\textit{MAE}) for random forest; (C) \textit{<MAE>} value when using different lengths of historical data, the length is determined by variable $n$. X-axis shows the number of days ($n$) we use as features. For example, $n=1$ means we choose to use the previous one day's features. }
    \label{pred_based_on_target_city}
\end{figure*}

\textit{Prediction based on historical disease data of target city}. People in the same city normally may have more contacts with each other, as they live, work and take the public transportation in the same city.
Thus the previously infected individuals may continue to spread the disease in the same city. This motivates us to explore only using the historical disease data of a target city as features to predict the number of infected in the future. Motivated by the fact that the number of infected is more correlated with the recent number of infected (as shown in Figure~\ref{lag tau autocorr}A), we choose to use the previous $n$ ($n \in [1, 7]$) days of infected as features to predict the number of infected at day $t$. In other words, if we use $I_{t}^{j}$ as the label, the features that are used for learning are $\{I_{t-n}^{j}, I_{t-n+1}^{j}, \cdots, I_{t-1}^{j}\}$. We show the results of using $n=1,2,\cdots, 7$ in Figure~\ref{pred_based_on_target_city}. Taking $n=1$ as an example, we use the number of infected at day $t-1$ as features and use the number of infected at day $t$ as labels for every city. After training the model
by using the disease data of the first 34 days (from January 25th to February 28th), we test the prediction performance on the remaining 10 days (from February 28th to March 9th) for every city. In Figure~\ref{pred_based_on_target_city}, we give the prediction performance of linear regression and random Forest, respectively. In Figure~\ref{pred_based_on_target_city}A and~\ref{pred_based_on_target_city}B, different curves indicate we use features from previous $n$ days as training data. The $x$ axis indicates the dates that we need to predict for the number of infected. The \textit{MAE} is relatively small for daily prediction except for March 2nd for both of the learning models.
Figure~\ref{pred_based_on_target_city}C shows the \textit{<MAE>} of the prediction over 10 days.
If we take a look at Figure~\ref{pred_based_on_target_city}A and~\ref{pred_based_on_target_city}C together, the smallest \textit{MAE} and \textit{<MAE>}
of linear regression are given by using historical disease data of $n=4$.  The prediction performance of random forest is shown in Figure~\ref{pred_based_on_target_city}B and~\ref{pred_based_on_target_city}C. The average \textit{MAE} decreases with the increase of $n$. When $n>1$, random forest shows better prediction performance than linear regression. Since linear regression and random forest both perform well when $n=4$, we list the regression coefficients of the features for the two algorithms in Figure~S8 in the Appendix. The most relevant feature is the number of infected in the previous step ($I_{t-1}$) for both algorithms, which indicates the markovian property of epidemic spreading on the population~\citep{zhan2018coupling}. But the other feature importance is quite different for the two algorithms.

\begin{figure*}[!ht]
\centering
	\includegraphics[width=12cm]{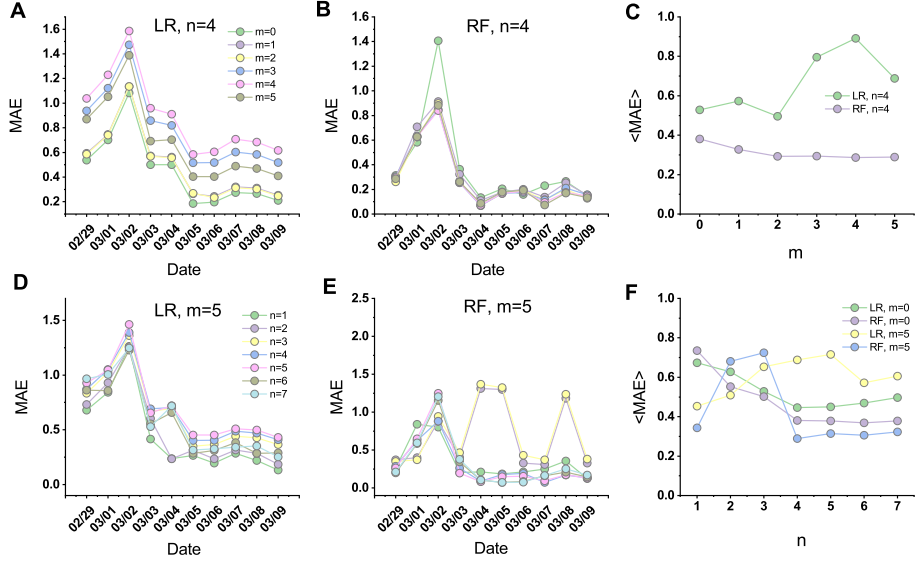}
	\caption{Prediction based on using the target city and nearby cities' historical disease data as features. (A) Daily prediction performance (\textit{MAE}) for linear regression, different curves corresponds to using different number ($m$) of nearby cities' historical data; (B) daily prediction performance (\textit{MAE}) for random forest, different curves correspond to using different number ($m$) of nearby cities' historical data; (C) \textit{<MAE>} value when using different number ($m$) of nearby cities' historical data as features. In (A-C), we fix $n=4$. (D) Daily prediction performance (\textit{MAE}) for linear regression, different curves corresponds to using different lengths ($n$) of historical data; (E) daily prediction performance (\textit{MAE}) for random forest, different curves corresponds to using different lengths ($n$) of historical data;(F) \textit{<MAE>} value when using different lengths ($n$) of historical data. }
    \label{pred_based_on_nei_city}
\end{figure*}

\textit{Prediction based on historical disease data of target city and nearby cities}.
In Figure~\ref{informationPCC}D, we showed that the provinces that are spatially close to Hubei tend to have more infected cases, which means the geographical distance between provinces (or cities) may have impact on the number of infected cases for a province (or city).
Thus, we include the historical data of nearby cities to predict the number of infected in a target city, besides the historical data of the target city. We use $m$ to
represent the number of nearby cities to be included as features.
In other words,
if we use $I_{t}^{j}$ as the label and choose $m=2$, the features that are used for learning are $\{I_{t-n}^{j}, I_{t-n+1}^{j}, \cdots,\linebreak I_{t-1}^{j}; N1(I_{t-n}^{j}), N1(I_{t-n+1}^{j}), \cdots, N1(I_{t-1}^{j}); N2(I_{t-n}^{j}), N2(I_{t-n+1}^{j}), \cdots,  N2(I_{t-1}^{j})\}$. In the features, $N1(*)$ and $N2(*)$ means features from the nearest city and the second nearest city, respectively. Similarly, if $m=3$, we use $N2(*)$ and $N3(*)$ to represent features from second and third nearest cities, respectively.
We first fix $n=4$ and test how the number of nearby cities' historical data will influence the prediction performance. We choose $m$ equals to $0, 1, 2, 3, 4, 5$, in which $m=0$ means we only consider the historical data of the target city as features.
The \textit{MAE} and \textit{<MAE>} for linear regression and random forest are shown in Figure~\ref{pred_based_on_nei_city}A-C. In Figure~\ref{pred_based_on_nei_city}A and~\ref{pred_based_on_nei_city}B, we show the prediction performance (\textit{MAE}) for everyday from February 29th to March 9th, 2020. In Figure~\ref{pred_based_on_nei_city}A and~\ref{pred_based_on_nei_city}C, we show that the \textit{MAE} and \textit{<MAE>} for linear regression. With the increase of $m$, the performance of linear regression first decreases until $m=2$. We obtain the lowest \textit{<MAE>} when $m=2$, which means including the nearby cities' infected number can help to increase the prediction performance for linear regression. We obtain similar results for random forest, as shown in Figure~\ref{pred_based_on_nei_city}B and~\ref{pred_based_on_nei_city}C.
 For the $m$ values we choose, we obtain the smallest \textit{MAE} and \textit{<MAE>} when $m=5$. In all the $m$ values we choose, random forest performs better than linear regression. In the following study, we fix $m=5$, i.e., we include the features from the first $5$ nearest cities in our model. We test how many previous days ($n$) we need to consider as features for prediction in Figure~\ref{pred_based_on_nei_city}D-F.
 Figure~\ref{pred_based_on_nei_city}D and~\ref{pred_based_on_nei_city}E show the daily prediction for linear regression and random forest, respectively.
 In Figure~\ref{pred_based_on_nei_city}F, we compare the \textit{<MAE>} for $m=0$ and $m=5$ with the change of $n$. When $m=5$, linear regression works better for small value of $n$, with the optimal \textit{<MAE>} achieved when $n=1$. When $n=1$ and $2$ for linear regression model, adding features from nearby cities helps to predict the number of infected, which are shown by the different between the curve of $m=0$ and $m=5$ in Figure~\ref{pred_based_on_nei_city}F. Random forest also works better when adding features from nearby cities for $n=1,4,5,6,7$, as shown in Figure~\ref{pred_based_on_nei_city}F.
 Additionally, random forest performs relatively better than linear regression for most $n$ values. Overall speaking, the optimal prediction is given by random forest when $m=5, n=4$.
 In a word,
we show from data (Figure~\ref{informationPCC}D) and model (Figure~\ref{pred_based_on_nei_city}) that if we want to predict \textit{COVID-19} or other contagious disease, we need to include the nearby cities' features in the model.
We further show the features that have top 10 largest feature coefficients when $m=5$ and $n=4$ for the two algorithms in Figure~S9 in the Appendix. Besides the features from the target city, the features from the nearby cities also contribute to the prediction of \textit{COVID-19}.
The top three most relevant features for linear regression are $I_{t-1}, N3(I_{t-1}), N3(I_{t-2})$. The top three most relevant features for random forest are $I_{t-1}, I_{t-2}$ and $I_{t-3}$. If we compare the most relevant features of only using the historical data of the target city (Figure~S8 in the Appendix), we find that random forest is more stable with regard to feature importance.

\subsection{Prediction based on disease and information data}
 In Figure~\ref{informationPCC} and S6 in the Appendix, we showed that information volume of a province is positively correlated with the population, GDP, GDP per captita, electricity consumption and highway transport volume. Particularly, provinces that are close to Hubei tend to have high information volume. Also, the number of infected in the provinces is highly
correlated to the information volume. Generally, the information related to a disease starts to spread after the disease occurs. Therefore, there is a time delay between the spread of the disease and disease information. We use $H_{t}^j$ to represent the information volume of city $j$ at time $t$. For city $j$, we have the daily infected time series, i.e., $I^j=\{I_1^j, I_2^j,\cdots, I_T^j \}$ and daily information time series, $H^j=\{H_1^j, H_2^j,\cdots, H_T^j \}$. We compute the time-delay ($\tau$) Pearson correlation between these two time series. For a specific value of $\tau$, we compute the Pearson correlation coefficients $r_{\tau}^j$ between the following two time series, i.e., $I^j(\tau+)=\{I_{\tau+1}^j, I_{\tau+2}^j,\cdots, I_T^j \}$ and $H^j(\tau-)=\{H_1^j, H_2^j,\cdots, H_{T-\tau}^j \}$.
The average correlation coefficient for all the cities is denoted as $r$ for a specific $\tau$. We show how $r$ changes with $\tau$ ranging from $0$ to $20$ in Figure~\ref{lag tau autocorr}B (blue curve). For each of the message, we also use natural language processing to classify the emotion of the messages, either positive, negative or neutral. Thus we give the correlation between the infected time series and positive, negative and neutral information time series as well in Figure~\ref{lag tau autocorr}B, as shown by the yellow, grey and green curves, respectively. For the four curves, the average correlation coefficient $r$ shows similar pattern.
The average correlation coefficient $r$ is always negative, with the lowest value $r=-0.39$ when $\tau=10$ when we use all the information time series for correlation calculation. The negative values of $r$ indicate the two time series are negatively correlated and also further suggest that information may have a delay effect on suppressing disease spreading.

\begin{figure*}[!ht]
\centering

	\includegraphics[width=6cm]{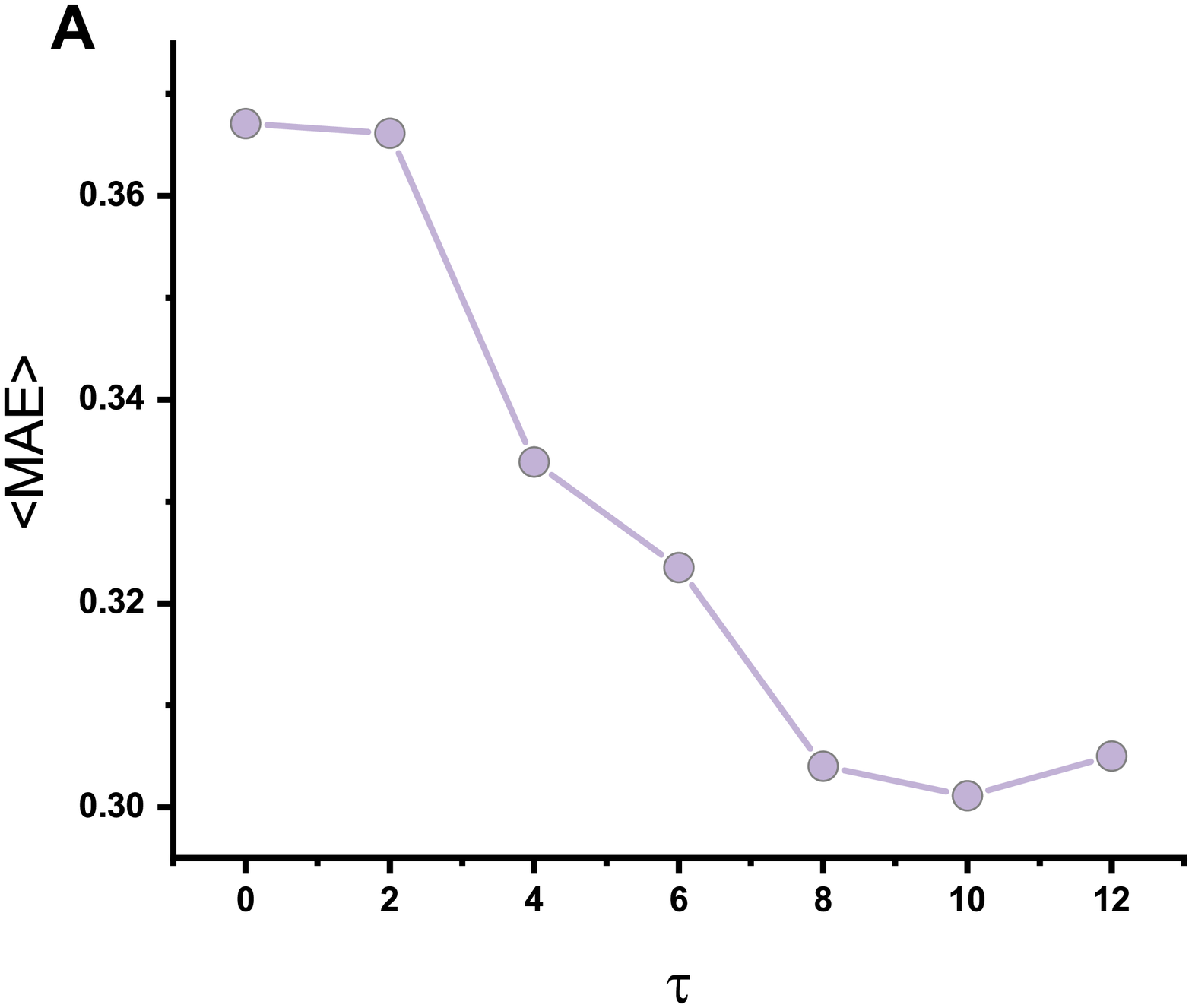}
	\includegraphics[width=6cm]{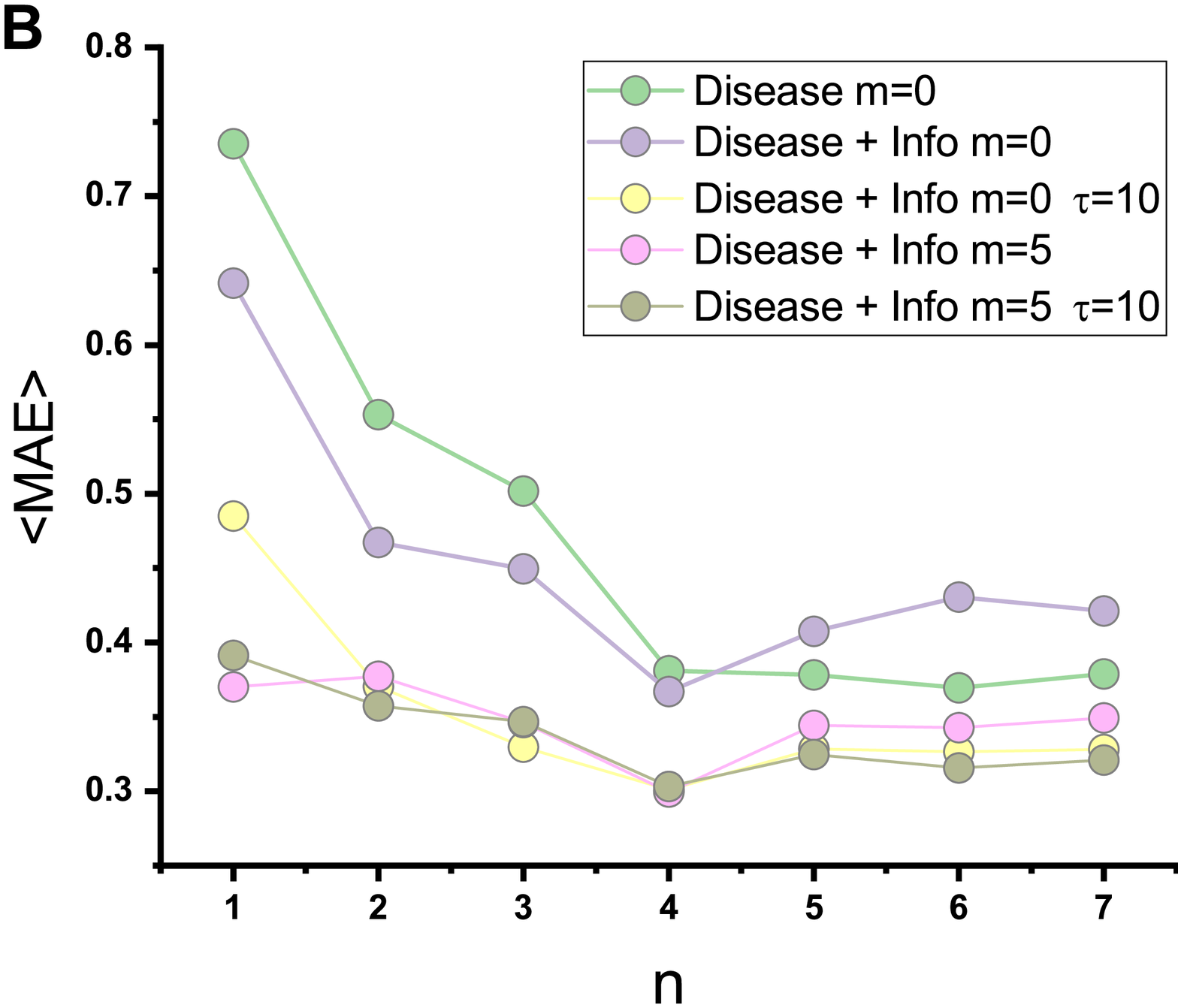}
	\caption{(A) Random Forest prediction based on disease and information data with $m=0, n=4$, we use $\tau$ to represent the delay of information as features for disease prediction. (B) Prediction based on using different groups of features: Black curve uses target city's disease data as features; Red curve uses target city's disease and information data as features; Blue curve uses target city's disease and information data as features. For information data, we use a time delay $\tau=10$; Purple curve uses target city and the top 5 nearest cities' disease and information data as features; Orange uses target city and the top 5 nearest cities' disease and information data as features. For information data, we use a time delay $\tau=10$.}
    \label{I Info lag tau corr}
\end{figure*}

Therefore, we further consider adding features derived from information data and explore how it is related to the prediction of \textit{COVID-19}.  As linear regression always shows worse performance than random forest when we use disease data as features, we will focus on using random forest for disease prediction in the following study.
For each city, we consider to use the daily information volume $H_t^j$ as additional features, besides the daily number of infected.  Figure~\ref{lag tau autocorr}B has shown that the correlation between disease and information series achieves the strongest negative value when $\tau=10$. We explore to use the delayed information data as features in our models.  We take $m=0, n=4$ as an example to illustrate how we use the disease and information data for prediction. The features that are used for learning are $\{I_{t-4}^{j},I_{t-3}^{j},I_{t-2}^{j},I_{t-1}^{j},H_{t-1-(\tau+3)}^{j},H_{t-1-(\tau+2)}^{j},H_{t-1-(\tau+1)}^{j},H_{t-1-\tau}^{j}\}$. We show \textit{<MAE>} of random forest for the prediction of 10 days in Figure~\ref{I Info lag tau corr}A. With the increase of delay $\tau$, \textit{<MAE>} decreases as well. We achieve the lowest \textit{<MAE>} when $\tau=10$. This means that using the information 10 days ago can help for the prediction of disease most, the result is consistent with the correlation between disease and information data we show in Figure~\ref{lag tau autocorr}B.

To further show how disease information can help for the prediction of disease, we compare the performance of using different groups of features in Figure~\ref{I Info lag tau corr}B. The first group of features are the ones based on the target city's historical disease data. The results are given by the grey curve with blue circles in the figure. The second group of features are based on the historical disease and information data of the target city. We use the grey curve with
yellow circles to represent in the figure. In the third group (grey curve with grey circles in the figure), we use the historical disease data and the information data of the target city as features, but we consider the information with a $\tau=10$ days delay. In the fourth group (grey curve with dark blue circles), we consider to use the target city and its top 5 nearest cities' disease and information data as features. In the fifth group (grey curve with pink circles), the only difference with that of fourth group is that we consider the information data has a $\tau=10$ days delay. If we compare the first three curves, we find that including the features from information can help to decrease the \textit{<MAE>}, i.e., increase the prediction performance. To be specific, the second group uses information features without delay, it can decrease \textit{<MAE>} for $n \in [1, 4]$. Particularly, the time-delayed information (grey curve with grey circles) can largely increase the prediction performance for all the values of $n$ compared to using information without delay. In the last two groups, we add disease and information features from the top five nearest cities. The last two groups show relatively better performance than the first three groups. The fifth group which uses delayed information features show better or sometimes similar performance compared to the fourth one which uses information without delay. The five groups of features we use indicate that both information and the features from nearby cities can help to increase the prediction performance. In addition, we find that each of the five curves shows lowest \textit{<MAE>} when $n=4$. We show the top 10 most relevant features for the second, third, fourth and fifth group when $n=4$ in Figure~19 in the Appendix, as we have already shown the feature coefficients of the first group in Figure~17 in the appendix. In all the five groups, they share the same top 2 most relevant features, i.e., $I_{t-1}$ and $I_{t-2}$. Particularly, the feature coefficient of $I_{t-1}$ is always larger than $0.8$, which means $I_{t-1}$ plays an essential role in predicting the disease. In Figure~S10, we also find that the features from information and nearby cities play important roles in predicting the disease. We note that we also
test on using the number of positive, negative or neutral messages as features in the algorithms. However, they show worse performance compared with using all the messages as features, so we haven't shown the curves in the Figure. Regarding to the content of the messages, we extracted Chinese words which have the same meaning with the following words: 'Policy', 'Social distancing', 'mask' and 'vaccine'. We count how many times these four kinds of words appear everyday and show how they are correlated with the infected time series with lag $\tau$, respectively. The correlation coefficients are given in Figure~\ref{lag tau autocorr}C and~\ref{lag tau autocorr}D, in which~\ref{lag tau autocorr}D shows the average correlation coefficients when $\tau \in [0, 20]$. We find these four types of words are almost negatively correlated with the infected time series when $\tau$ changes (except small $\tau$ values). That is to say, these four words are influential to control disease, especially for words that have the same meaning as 'vaccine' and 'policy'. Additionally, we also test on using the data from January 25th to February 21st for training and test. That is to day, we use the first 21 days as training set and the other 7 days as test set. The results are given in Figure S11 in the Appendix. We find similar results, i.e., both the features from nearby cities and information are helpful for improving the prediction accuracy.

\section{Discussion and Conclusion}
In this paper, we worked on the co-evolution processes of the spreading of COVID-19 and its information in China. The information data of COVID-19 came from well-known Chinese online social medium, such as Sina Weibo, authorized mobile news apps, articles on Wechat’s official accounts, etc. Therefore, the data is quite representative with regard to the public reaction to such a severe contagious disease. We visualized how the disease and information are spatially localized in China via using the province-level data.  We find that there is a delay between the spread of COVID-19 and its information, i.e., the peak of the disease is at the beginning of February, 2020, whereas the peak of information is around the beginning of March, 2020. Additionally, we researched on how the information data, i.e., the number of messages, in each province is correlated with the disease data and other variables. The correlation shows a province with more infected numbers tends to be more mentioned online. Also, both information and disease shows negative correlation with the distance to Hubei. That is to say, the population tends to have more attention on the disease when it spatially comes to their neighborhood. We also showed weekly and hourly pattern of the messages in each platform in the Appendix.

Based on the analysis of the disease and information time series, we proposed to use different characteristics to predict the number of infected in the city-level, including 322 cities in China which have nonzero infected cases. We use two machine learning algorithms, i.e., linear regression and random forest, to train and predict the number of infected. Meanwhile, the features that are used in the algorithms can be classified into three categories: (\romannumeral1) we use the historical number of infected in each target city as the features; (\romannumeral2) Besides (\romannumeral1), we also include the historical number of infected of the target cities' neighboring cities; (\romannumeral3) Besides (\romannumeral1) and (\romannumeral2), we include the number of messages in each target city and its neighboring cities as features. We find that random forest always performs better than linear regression when we use different categories of features. Including the features from neighboring cities, i.e., the cases of (\romannumeral2) and (\romannumeral3), improves the prediction performance. This is consistent with the correlation analysis given at the province level, where we find both the distance and information is highly correlated with the number of infected.

The use of online social  media enables us to make the observational analysis between COVID-19 and its information, and thus explore whether information can be useful for the prediction of disease. The results shown in this work may suggest a new way of predicting emerging infections. However, we claim there are some limitations which is beyond the analysis of this work. We only explore the COVID-19 data in China, the data from the rest of the world is not included. We deem that it should be very interesting to explore the co-evolution of these two processes in the other countries, especially in the countries that have several waves of epidemic. As a matter of fact, there are still epidemic outbreaks in the other countries where people actually are tired against the disease. The multiple waves of epidemic~\citep{pullano2021underdetection, aleta2020modelling,liu2020markov,zhan2016roles} and the public's tiredness against the disease may induce different co-evolution phenomena between these two spreading processes. Additionally, we only consider using the number of messages as features for the prediction, the reliability of the information is beyond our analysis. We claim that it is also an interesting direction to explore whether we need to exclude the fake news in the prediction~\citep{loomba2021measuring}.

\section{Method and Data Description}

\subsection{Data description}

\emph{Information data.}
We collect the information data of \textit{COVID-19} from Qingbo Bigdata (\url{www.gsdata.cn}) based on the Chinese words that are in the medical discourse about \textit{COVID-19}.  The key words that we select are  'wuhan', 'feiyan', 'xinguan', 'yiqing'.
We track the omnimedia information from January 25th, 2020 to March 24th, 2020, including Sina Weibo (\url{http://weibo.com/}, China’s largest micro-blogging system), authorized mobile news apps, articles on \textit{Wechat}’s official accounts (China’s most popular instant message app), ordinary news webpages, newspapers as well as various online forums. We also record the location (city) of information source. And there are totally 322 cities in our dataset (we exclude the cities that have zero disease infected cases), including $20$ county-level cities. In total, we have collected $163.1$ million of messages, in which $154.8$ million messages contain both time and geographic information. The data shows how people reflect on the pandemic evolves with time. The detailed analysis of the data on each of the platforms is given in Section 1.1 in the Appendix.

\emph{Disease and other data.} The data of daily infected cases of \textit{COVID-19} for every city is from China Centers for Disease Control and Prevention (CDC). We collect data from January 25th, 2020 to March 24th, 2020. We only consider the cities that have infected cases, i.e., the cities that have  zero infected cases are excluded, resulting in daily infected data for $322$ cities. The resident population data is collected from 'China Statistical Yearbook 2019'.
\subsection{Linear regression}
Linear regression (LR) is a regression algorithm that aims to study the linear relationships between variables. Given a data set $\mathcal{D}_{n} = \{y_{i}, x_{i1}, x_{i2}, \cdots, x_{im}\}_{i=1}^{n}$ with $n$ training samples, $x_{i} = (x_{i1}, x_{i2}, \cdots, x_{im})$ are treated as features and $y_{i}$ is the training label. We use linear regression to train the data. The objective of linear regression is as follows:
\begin{equation}
\centering
\min   \sum_{i=1}^n(y_{i}-\sum_{j=1}^mx_{ij}\beta_{j})^2, \label{eq1}
\end{equation}
where $n$ is the number of training samples, $m$ is the dimension of each $x_i$, $\beta_{j}$ is the regression parameter we need to learn.

\subsection{Random forest}
    Random forest (RF) is a learning algorithm that builds an ensemble of decision trees and merges them together to get an accurate and stable prediction. Each decision tree in a random forest is trained on a data point sample, which is randomly chosen from the rows and columns of training data. Therefore, we consider a random set of features for splitting at each decision node. Random forest has the advantage of learning the non-linear relationship between features and labels. Also, random forest can help to avoid overfitting by combining many decision trees with randomness. It uses bagging in the training process. Given a data set $\mathcal{D}_{n} = \{y_{i}, x_{i1}, x_{i2}, \cdots, x_{im}\}_{i=1}^{n}$ with $n$ training samples, $x_{i} = (x_{i1}, x_{i2}, \cdots, x_{im})$ are treated as features and $y_{i}$ is the training label. The data point sample used for fitting in each decision tree is generated by repeatedly selecting a random sample with replacement of $\mathcal{D}_{n}$. Suppose the number of trees we consider is $B$, for each decision tree $b \in [1, B]$, we do the following two steps:
\begin{itemize}
\item We randomly do the sampling with replacement on $\mathcal{D}_{n}$ to get $n$ training examples, we name them as $\mathcal{D}_{n}^{b}$. Data $\mathcal{D}_{n}^{b}$ is the a data point sample we mentioned previously.
\item We train decision tree $b$ on $\mathcal{D}_{n}^{b}$.
\end{itemize}

When we do the prediction, the prediction on the test set is the average over all the decision trees. The bagging procedure can decrease the variance of the model without increasing the bias, which can lead to better prediction performance.



\section*{Acknowledgments}
This study was partially supported by Zhejiang Provincial Natural Science Foundation of China (Grant No. LR18A050001), XPCC's Key Scientific and Technology Project (No.2021AB034), the National Natural Science Foundation of China (Grant Nos. 61873080, 61673151), the Major Project (Grant Nos. 19ZDA324, 19ZDA325) and the Key Project of The National Social Science Fund of China (Grant No. 19BXW084).

\bibliographystyle{elsarticle-num}
\bibliography{sample}

\end{document}